\documentclass[wssquare]{ws-procs975x65}

\newcommand\neededonlyforarXiv[2]{#1}

\def\lsim{\mathop{\hbox{${\lower3.8pt\hbox{$<$}}\atop{\raise0.2pt\hbox{$\sim$}}
$}}} 
\def\gsim{\mathop{\hbox{${\lower3.8pt\hbox{$>$}}\atop{\raise0.2pt\hbox{$
\sim$}}$}}}

\newcommand\hMpc{\mbox{$h^{-1}$ Mpc}}

\providecommand\hMpc{\mbox{$h^{-1}$ Mpc}}

\providecommand\apjl{Astrophys.~J.~Lett.}                 
\providecommand\apj{Astrophys.~J.}                 
\providecommand\prd{Phys.~Rev.~D}
\providecommand\mnras{Mon.~Not.~Roy.~Astr.~Soc.}
\providecommand\aap{Astron.~Astroph.}

\providecommand\prd{Physical Review D}

\providecommand\apjl{Astrophys.J.Lett.}                 

\newcommand{\D}{\mathcal{D}}
\newcommand{\Dn}{\mathcal{D}_n}
\newcommand{\C}{\mathcal{C}}
\newcommand{\aD}{a_\mathcal{D}}

\newcommand{\ddaD}{\ddot a_\mathcal{D}}

\newcommand{\QD}{Q_\mathcal{D}}
\newcommand{\QC}{Q_\mathcal{C}}

\newcommand{\vgrad}{v_{i,j}}

\newcommand{\ainvIC}{\langle\mathbf{I}\rangle_{\C}}
\newcommand{\ainvIIC}{\langle\mathbf{II}\rangle_{\C}}

\newcommand{\omm}{\Omega_{\mathrm{m}}}
\newcommand{\omlam}{\Omega_{\Lambda}}

\neededonlyforarXiv{
  \providecommand\eprint[1]{\href{http://arXiv.org/abs/#1}{arXiv:#1}} 

  \usepackage{hyperref}
  \hypersetup{
    colorlinks=true,
    citecolor=blue,
    linkcolor=magenta,
    filecolor=cyan,      
    urlcolor=cyan,
  }
}{
  \providecommand\eprint[1]{{\tt [arXiv:#1]}}
  \providecommand\href[2]{\url{#1}} 
}

\begin{document}
\hypersetup{pdfpagelabels=false,breaklinks=false,linktocpage}



\newcommand\figQzeroTestInvCompare{
  \begin{figure}
    \vspace{3ex}
    \begin{center}
      \scalebox{1}{\input{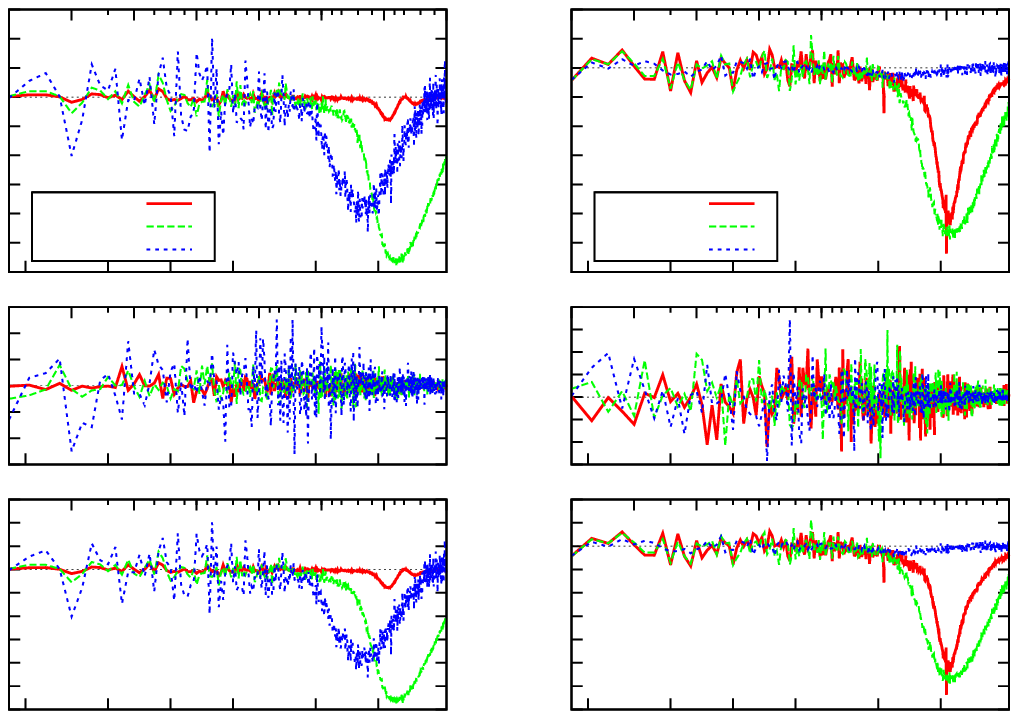}}
      \caption{$Q_{\mathcal{C}} = 0$ test:
        upper panel -- $\QC$, middle -- $\ainvIC$, bottom -- $\ainvIIC$, 
        left panels: VC~EdS simulation, right panels:~Gadget-2~EdS simulation,
        calculated for $z=\{10,3,0\}$. The source of strange negative values for $\QC$
        is $\ainvIIC$. Calculations done using default averaging method in DTFE i.e.
        sampling over delaunay cells with approx. 100 samples per grid cell.
      }
      \label{inv}
    \end{center}
  \vspace{-2ex}
  \end{figure}
}

\newcommand\figQzeroTestMcompare{
  \begin{figure}
    \vspace{-2ex}
    \begin{center}
      \scalebox{1.0}{\input{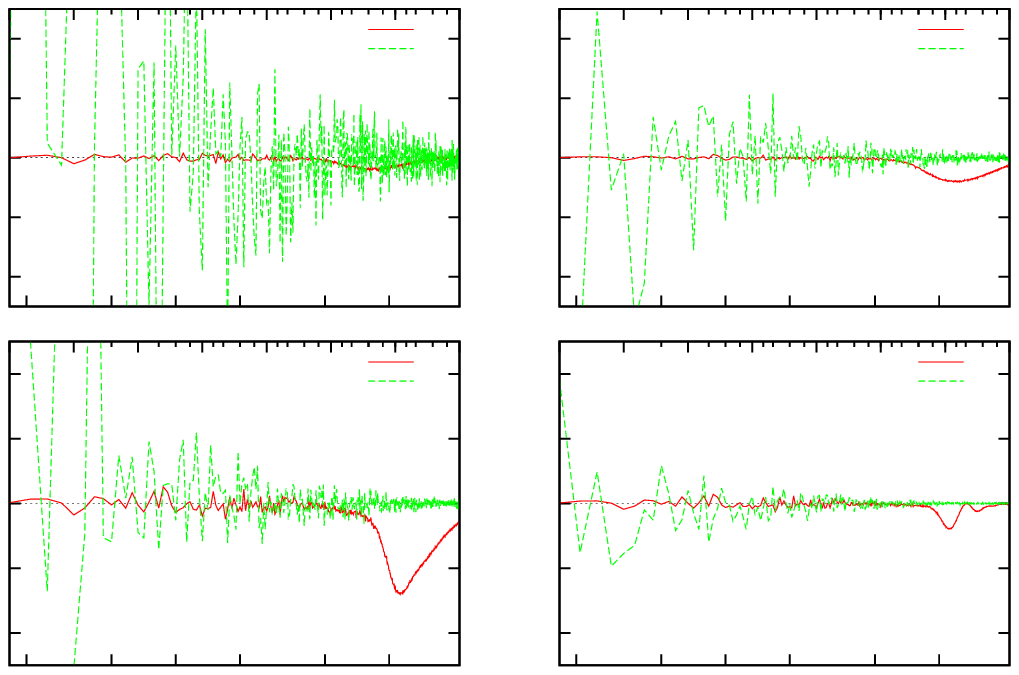}}
      \caption{Comparison of two interpolation methods: sampling over DT cells (red, solid line) and sampling
        over grid cells (green, dashed line) for $z=0$ (left top pane), $z=1.5$ (right top), $z=5$ (left down)
        and $z=10$ (right down).}
      \label{VirgoEdS_m2_four_z}
    \end{center}
  \end{figure}
}

\newcommand\figPDFs{
  \begin{figure}
    \begin{minipage}[b]{0.5\linewidth}
    \centering
      \scalebox{0.5}{\input{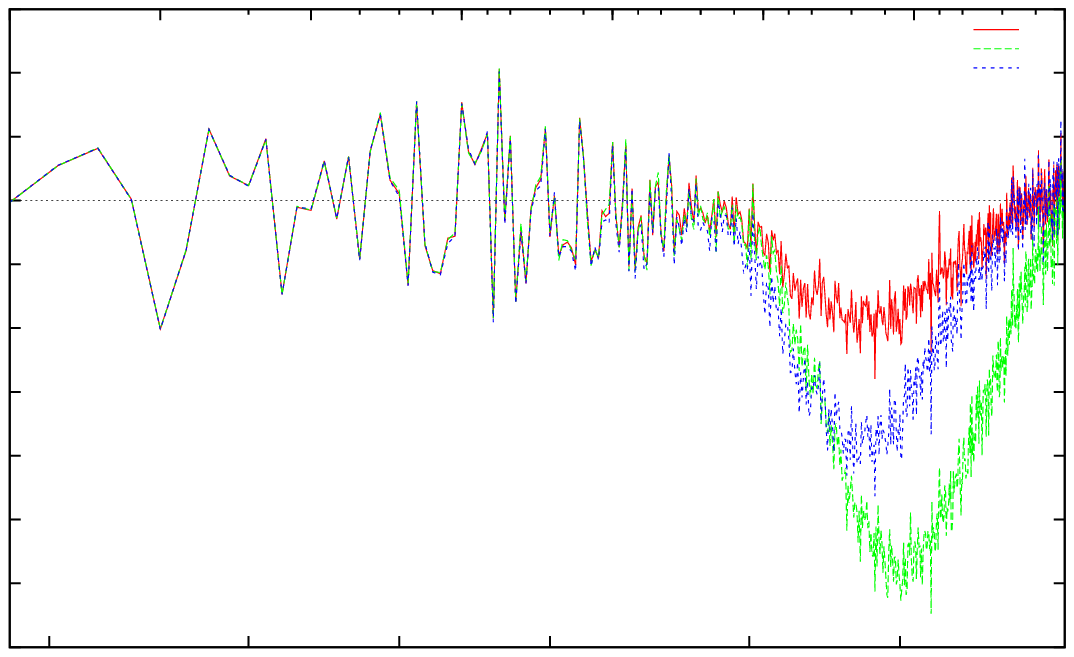}}
      \caption{DT interpolation method (sampling in Delaunay cells) as a function
        of sample size for VC~EdS, $z=0$:
      50 samples --- green, 100 --- blue (default), 200 --- red.}
      \label{VirgoEdS_m1_sampling_compare}
    \end{minipage}
    \hfill
    \begin{minipage}[b]{0.45\linewidth}
      \centering
      \newcommand{\mylen}{\textwidth}
      \includegraphics[width=\mylen]{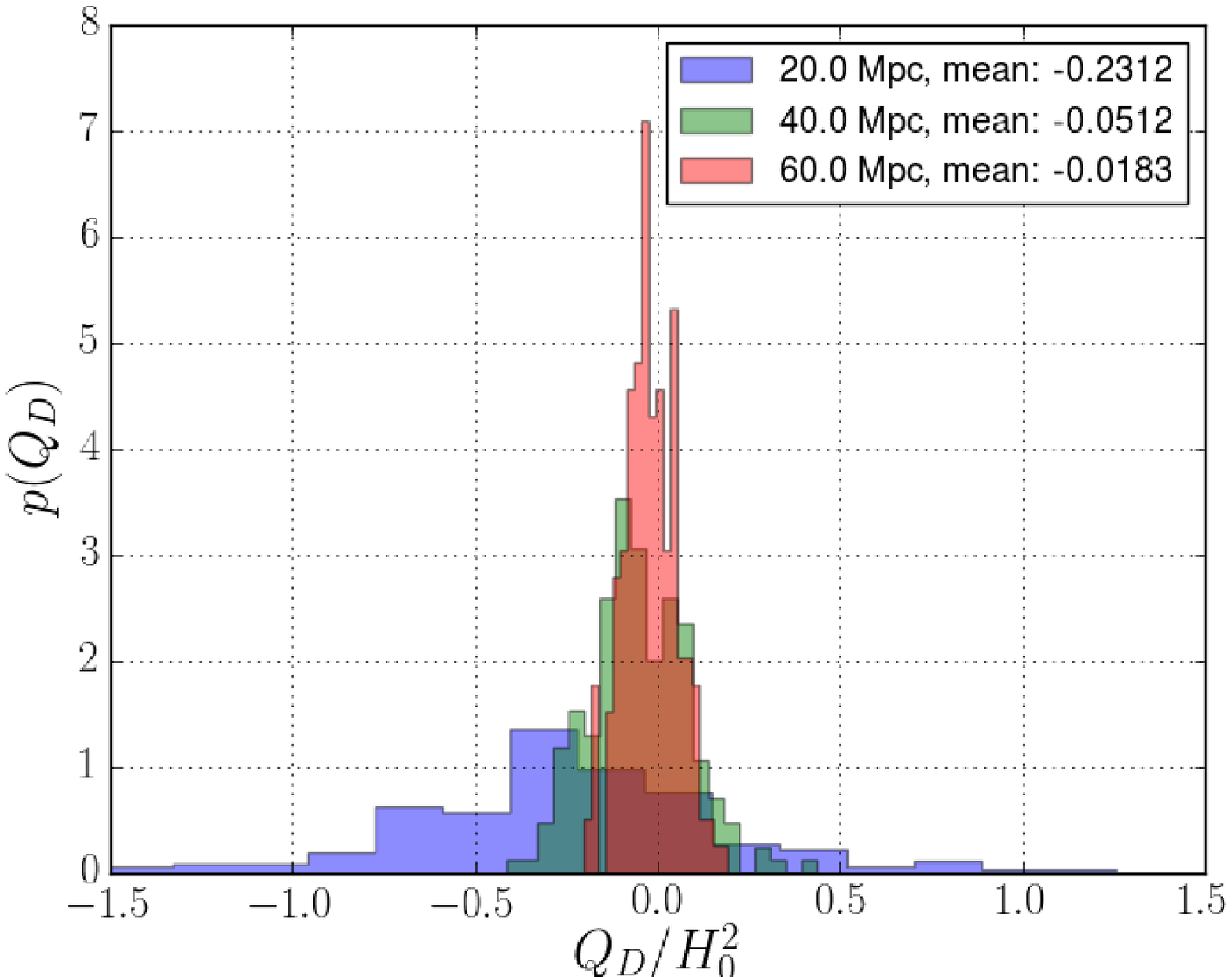}
      \caption{Histogram of $\QD$ 
        from sample of 200 times randomly chosen domain $\D$ of three different
        sizes: $20\hMpc$, $40\hMpc$ and $60\hMpc$ divided by $H^2_0$ in order to have
        dimensionless units.}\label{PDFs} 
    \end{minipage}
  \end{figure}
}



\markboth{Kazimierczak}
         {Kinematical backreaction $Q$ with DTFE}
\title{Newtonian kinematical backreaction in cosmological\\
  $N$-body simulations with Delaunay Tesselation:\\
  ``zero test'' and scale dependence}

\author{Tomasz~A.~Kazimierczak} 

\address{Toru\'n Centre for Astronomy, 
  Faculty of Physics, Astronomy and Informatics,
  Grudziadzka 5,
  Nicolaus Copernicus University,
  ul. Gagarina 11, 
  87-100 Toru\'n, Poland}

\begin{abstract}
  
  The backreaction of inhomogeneities describes the effect of
  inhomogeneous structure on average properties of the Universe.
  We investigate this approach by testing the consistency of cosmological
  $N$-body simulations as non-linear structure evolves.
  Using the Delaunay Tessellation Field Estimator  
  (DTFE), we calculate the kinematical backreaction $Q$
  from simulations on different scales 
  in order to measure how much $N$-body simulations should be corrected for this effect.
  This is the first step towards creating fully relativistic and inhomogeneous $N$-body simulations.
  In this paper we 
  compare the interpolation techniques available in DTFE and
  illustrate the statistical dependence of $Q$ as a function of length scale.
\end{abstract}


\bodymatter


\section{Introduction}

Inhomogeneous structure of the Universe at scales below $500\hMpc$ is an undeniable fact.
The standard approach to model it (i.e., in the standard $\Lambda$CDM model approach) is to
perturb the homogeneous solution of the Einstein equations, i.e. the
Friedmann--Lema\'itre--Robertson--Walker (FLRW) metric.
However, there are strong suggestions that this approach may not be the best one especially 
for the late times (i.e. small redshifts $z\leq 3$ ) of the Universe evolution
(e.g., Ref.~\refcite{deLappGH86}).

Another approach, \emph{scalar averaging} using general relativity (GR), 
introduces kinematical and curvature \emph{backreaction} of the inhomogeneous 
structure of the Universe, in principle without assuming a homogeneous background.
Here we focus only on the Newtonian version of this approach \cite{BKS00,BuchertEhlers97} since current $N$-body simulations
are Newtonian (within the expanding FLRW background, which is rigid in comoving coordinates). Our motivation is to investigate 
whether the effect,
namely kinematical backreaction $Q$, described in Ref.~\refcite{BKS00}, exists in $N$-body simulations as the model predicts, 
and check how big this effect is compared to that expected analytically.

\section{Method}

In the Newtonian approach there is only the kinematical backreaction\footnote{Curvature backreaction is 
obviously zero because of the flat Euclidean spatial section.} $Q$ which occurs in the
generalised acceleration law:\cite{BuchertEhlers97,BKS00}
\begin{equation}
3\frac{\ddaD}{\aD} + 4\pi G \left< \rho \right>_\mathcal{D} - \Lambda = \QD \label{gen_expansion_law},
\end{equation}
where
\begin{equation}
  \QD := \frac{2}{3} \left( \langle \theta^2\rangle_\mathcal{D} - \langle \theta\rangle^2_\mathcal{D} \right)
  + 2 \langle\omega^2 - \sigma^2 \rangle_\mathcal{D} \label{Q_defn}. 
\end{equation}
The volume-weighted average 
$\langle \cdot \rangle_\mathcal{D}$ is defined as 
a volume integral normalised by the volume of the domain $\D$ for which averaging is performed.
The quantities $\theta$, $\omega$ and $\sigma$ 
are respectively the expansion rate and the rates of vorticity and shear, defined in the standard
way as elements of the decomposition
of the velocity gradient tensor $v_{i,j}$ in three parts: the trace $\theta$, the symmetric
part $\sigma_{ij}$ and the antisymmetric 
part $\omega_{ij}$.
Equation \ref{Q_defn} can be rewritten in terms of the tensor invariants:
\begin{equation}
  \QD = 2 \langle \mathbf{II}(v_{i,j})\rangle_\mathcal{D}
  - \frac{2}{3} \langle \mathbf{I}(v_{i,j}) \rangle^2_\mathcal{D},\label{Q_inv}
\end{equation}
where
\begin{subequations}
  \label{inv_defn}
  \begin{align}
    \mathbf{I}(v_{i,j})  &= \mathrm{tr}(v_{i,j}) = \theta, \label{invI} \\
    \mathbf{II}(v_{i,j}) &= (\mathrm{tr}(v_{i,j}))^2-\mathrm{tr}((v_{i,j})^2)=
    \omega^2-\sigma^2+\frac{1}{3}\theta^2\label{invII}
  \end{align}
\end{subequations}
and $\vgrad := \partial_j v_i$.

\subsection{Interpolating and averaging fields from $N$-body simulations}
In order to calculate $Q$ from $N$-body simulations for a given domain $\D$ one needs to interpolate
the velocity field (and its gradient)
from a discrete set of points, since in $N$-body simulations all information about fields is encoded
in particles.
There are several methods for interpolating fields from a set of points
(e.g. SPH, CIC). We chose the \emph{Delaunay Tessellation Field Estimator}%
\footnote{
  The code is free-licensed and available at: \url{http://www.astro.rug.nl/~voronoi/DTFE/dtfe.html}.
  DTFE uses CGAL---the Computational Geometry Algorithms Library, also free-licensed
  (\url{http://www.cgal.org/}).
}
(DTFE) \cite{DTFE00} method, which is based on
Delaunay Triangulation (DT).\cite{Delaunay1934} The advantage of this choice is that it interpolates
the velocity field and its gradients in a very natural way, i.e. by linear interpolation inside every Delaunay
cell (a tetrahedron in the 3D case).\cite{BernardeauWeygaert96} Cells are constructed from a discrete
set of points
(particles) in such a way that every tetrahedron constructed from 4 particles fulfills the requirement that
there is no other particle inside a sphere circumscribed on that tetrahedron.

\subsection{ ``$\QC=0$'' test for periodic boundary condition} \label{s-Q-zero-test}
From \eqref{Q_defn} and \eqref{gen_expansion_law} it is clear that the backreaction $Q$ in the Newtonian
approach is purely 
of a kinematical origin. Moreover, because of the way in which it is defined, it will always be zero if
there is no boundary,
e.g. if the ``boundaries'' are periodically identified, as is the case for cosmological 
numerical simulations.\footnote{The problem of boundaries in cosmological simulations is ``solved''
by setting up 
a $T^3$ topology of the simulation box, i.e. periodic translation of the fields through the opposite
faces of the simulation 
box.}. This is due to the fact that, by using Gauss' theorem, $\QD$ in comoving coordinates
can be expressed as an integral over the surface $\partial\D$ (see eq.~(10) of \cite{BKS00}).
This gives the possibility to test existing simulations for consistency if they preserve $\QC=0$
($\mathcal{C}$ denotes the whole simulation domain
with $T^3$ topology, ``periodic boundary conditions''). 
If one divides the whole simulation box domain onto $N^3$ smaller cubes (or domains of another shape) 
of equal volume $|\mathcal{D}_n| = |\mathcal{C}|/N^3$, then the following equation will be valid:
\begin{equation}
  \QC = 2\langle\mathbf{II}\rangle_\mathcal{C} - \frac{2}{3} \langle\mathbf{I}\rangle_\mathcal{C}^2
  = \frac{2}{N^3} \sum_{n=1}^{N^3} \mathbf{II}_{\Dn}
- \frac{2}{3} \left(\frac{1}{N^3}\sum_{n=1}^{N^3} \mathbf{I}_{\Dn}\right)^2
= 0 \label{Q_zero} 
\end{equation}
for $N>0$. 


\section{Results}

For calculations we used Einstein--de~Sitter 
($\omm=1, \omlam=0$) $N$-body simulations:
(i) a Virgo Consortium (VC) simulation from \cite{Jenkins98virgo} (simulation SCDM1, hereafter VC~EdS)
and (ii) our own simulation performed with Gadget-2 (hereafter, Gadget-2~EdS)
with the same box size and number of particles and cosmological parameters as in VC~EdS 
(simulation box size: $240\hMpc$, $256^3$ particles, $h=0.5$).

We tested vanishing $\QC$ with periodic boundary conditions (the ``$\QC=0$'' test)
with $4 \le N \le 512$. For any $N$, this sets up the grid resolution which corresponds to  
sub-box domains $\Dn$ of a fixed size ($n=1,\ldots,N$); and probes all ranges of scales
from $L_\mathcal{C}/512$
to $L_\mathcal{C}/4$ for a simulation with a box side length $L_\mathcal{C}$. 
The velocity gradient was calculated using DTFE with the
default volume averaging method, i.e. Monte Carlo sampling
over DT cells with approximately 100 samples\footnote{The number of samples depends
on the ratio between the DT cell volume and the chosen grid cell volume.} per each grid cell
for a given estimate.

Figure~\ref{inv} shows $\QC$ as a function of the size $L_{\Dn} = L_\mathcal{C}/N$ of sub-domains
$\Dn$ (eq.~\ref{Q_zero}),
or equivalently, as a function of $N$.

  \begin{figure}
    \vspace{3ex}
    \begin{center}
      \scalebox{1}{\input{QD_zero_test_inv_compare}}
      \caption{$Q_{\mathcal{C}} = 0$ test:
        upper panel -- $\QC$, middle -- $\ainvIC$, bottom -- $\ainvIIC$, 
        left panels: VC~EdS simulation, right panels:~Gadget-2~EdS simulation,
        calculated for $z=\{10,3,0\}$. The source of strange negative values for $\QC$
        is $\ainvIIC$. Calculations done using default averaging method in DTFE i.e.
        sampling over delaunay cells with approx. 100 samples per grid cell.
      }
      \label{inv}
    \end{center}
  \vspace{-2ex}
  \end{figure}

$\QC$ does not stay close to zero for every $N$, particularly for $N$ 
ranging from 128 to 512, where $\QC$ is negative for both the VC EdS and Gadget-2 EdS simulations
(top panels of fig.\ref{inv}).
This corresponds to $2\hMpc \gsim \Dn \gsim  0.5 \hMpc$. 
Moreover fig.~\ref{inv} shows $\ainvIC$ (middle panels) and $\ainvIIC$ (bottom panels) from which---with
respect
to the $\Dn$---$\QC$ was calculated. It is clear that negative values of $\QC$ comes
from $\ainvIIC$. Equation~\eqref{invII} shows that this significant deviation from zero values comes
either from overestimation of $\sigma$ or underestimation of $\omega$ and/or $\theta$,
since the square of the former provides a negative contribution
and square of the latter two positive input to $\ainvIIC$ (see eq.~\ref{invII}).

For a regular grid such as the one used here, 
DTFE has two built-in methods of interpolating fields to a grid location followed
by a local averaging procedure (hereafter, ``averaged interpolation''):
(i) sampling randomly over Delaunay cells
(hereafter, the DT method; as in fig.~\ref{inv}); or (ii) sampling (randomly or not) within the grid cell
(hereafter, the grid method).
We have performed the same test for both averaged interpolation methods, and varied the numbers
of random samples. 

Figure~\ref{VirgoEdS_m2_four_z} compares these two methods for 
the ``$\QC = 0$'' test with the VC EdS simulation.

  \begin{figure}
    \vspace{-2ex}
    \begin{center}
      \scalebox{1.0}{\input{QD_zero_test_VirgoEdS_m1_vs_m2_four_z}}
      \caption{Comparison of two interpolation methods: sampling over DT cells (red, solid line) and sampling
        over grid cells (green, dashed line) for $z=0$ (left top pane), $z=1.5$ (right top), $z=5$ (left down)
        and $z=10$ (right down).}
      \label{VirgoEdS_m2_four_z}
    \end{center}
  \end{figure}

The DT method (solid red line) is less noisy in general, but produces strong negative values
for $128 \le N \le 512$. Grid sampling (dashed green line) is
more noisy (especially for high $z$) but does not produce this feature for $N\gsim 120$,
even though the sample size per grid cell (by default, 20 random samples) is lower than for the DT method.
When $N$ is high, the number of particles per grid cell is low, so
that Delaunay cells will be large compared to a grid cell;
thus, a systematic error in volume-weighted averaging of the velocity gradient
could occur by sampling within the local Delaunay cells rather than within the grid cell.
This may explain the DT-method high-$N$ negative $\QC$ values, in which case
increasing the sample size should weaken the negative $\QC$---figure~\ref{VirgoEdS_m1_sampling_compare}
supports this.

\subsection{Statistics of $Q$---early results}

Motivated by the $\QC=0$ analysis, 
the DT method, with random sample size increased to 300 and grid
size $N=100$, was used to estimate 
probability density functions (PDF) of $\QD$ 
for the VC~EdS simulation at redshift $z=0$.
Figure~\ref{PDFs} shows PDFs for three different domain $\D$ sizes, $L_{\D} = 20, 40, 60\hMpc$;
in each case, 200 domains were chosen randomly. 
There is clear evidence of statistical scale dependence for $\QD$:
the smaller the domain $\D$, the more negative $Q$ tends to be 
(fig.~\ref{PDFs}).
More detailed calculations will be published soon.

  \begin{figure}
    \begin{minipage}[b]{0.5\linewidth}
    \centering
      \scalebox{0.5}{\input{QD_zero_test_VirgoEdS_z0m1_sampling_compare}}
      \caption{DT interpolation method (sampling in Delaunay cells) as a function
        of sample size for VC~EdS, $z=0$:
      50 samples --- green, 100 --- blue (default), 200 --- red.}
      \label{VirgoEdS_m1_sampling_compare}
    \end{minipage}
    \hfill
    \begin{minipage}[b]{0.45\linewidth}
      \centering
      \newcommand{\mylen}{\textwidth}
      \includegraphics[width=\mylen]{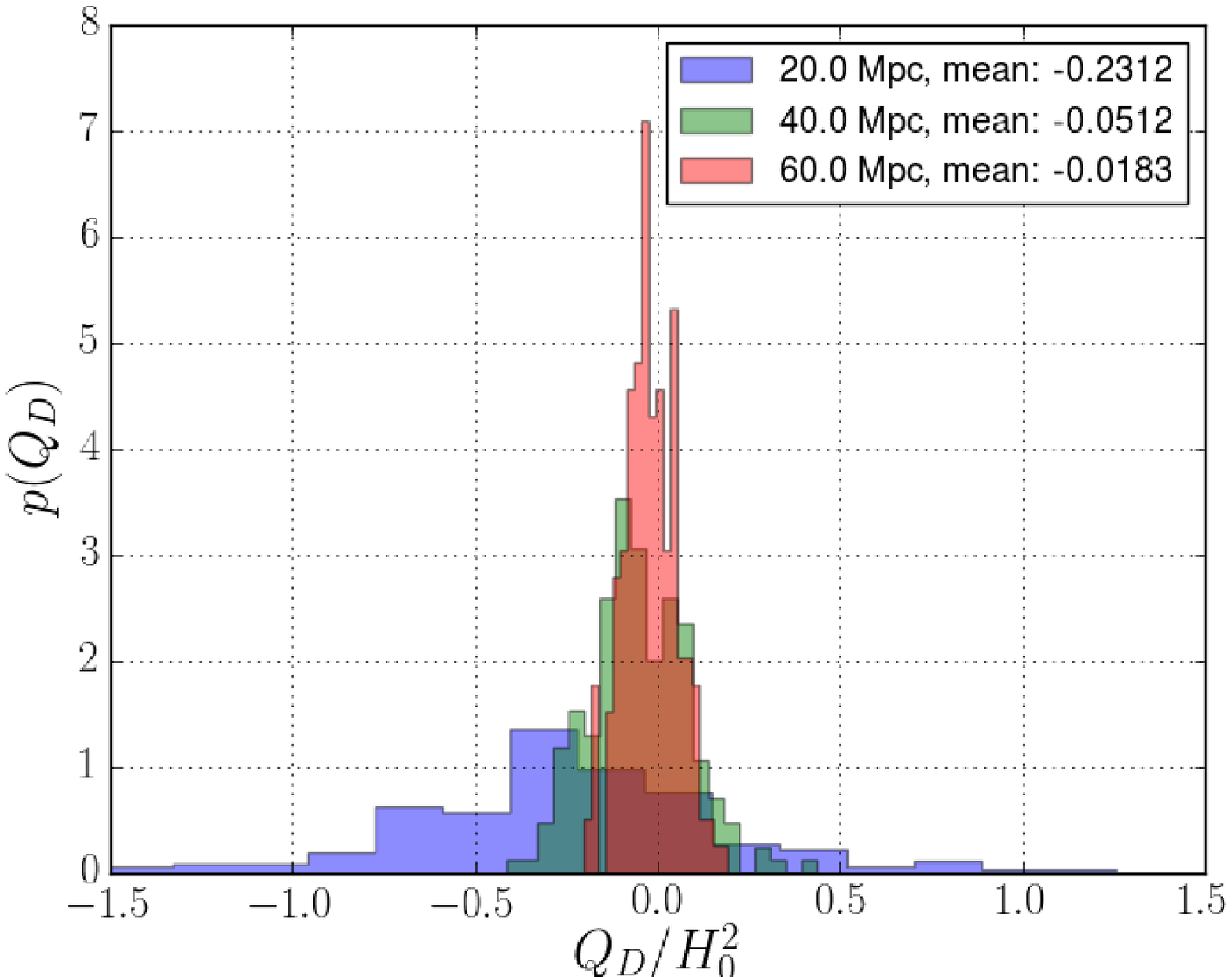}
      \caption{Histogram of $\QD$ 
        from sample of 200 times randomly chosen domain $\D$ of three different
        sizes: $20\hMpc$, $40\hMpc$ and $60\hMpc$ divided by $H^2_0$ in order to have
        dimensionless units.}\label{PDFs} 
    \end{minipage}
  \end{figure}

\section{Summary}

Using DTFE to perform the ``$\QC=0$ test'' for the VC~EdS and Gadget-2~EdS simulations suggests that 
the DT method (sampling over DT cells) introduces artificial behaviour for small grid cells, i.e.
when the number of grid cells is comparable to the number of
particles in simulation. At early epochs, 
the magnitude of the effect is comparable to that of the noise 
of the grid-sampling averaged interpolation method for big grid cells (low numbers of grid cells).
This systematic error can be reduced by increasing the number of random samples 
used for averaged interpolations within Delaunay tetrahedra
(fig.~\ref{VirgoEdS_m1_sampling_compare}),  
at the cost of slowing the calculation. The code could be improved by
implementing an exact calculation of volume-weighted averages in each grid cell 
(i.e. averaging the interpolation without random sampling).
The PDFs of $Q$ show statistical scale dependence (fig.~\ref{PDFs}).

\section*{Acknowledgments}

TK thanks the National Science Centre, Poland for partial support from an ETIUDA-2 grant
as well as to J.J.~Ostrowski, B.F.~Roukema and T.~Buchert for useful comments.
Some of this work was done under the
National Science Centre, Poland grant 2014/13/B/ST9/00845.

\end{document}